**Heavy Electrons**

# Sleuthing Hidden Order

Authors:

**V. Tripathi**
Department of Theoretical Physics
Tata Institute of Fundamental Research
Mumbai 400005, India

**P. Chandra and P. Coleman**
Center for Materials Theory
Rutgers, The State University of New Jersey
Piscataway, NJ 08854 USA

*Physicists have long debated whether the hidden order in $URu_2Si_2$ is itinerant or localized, and it remains inaccessible to direct external probes. Recent observation of an overdamped collective mode in this material appears to resolve this outstanding issue.*

As physicists probe the universe on different scales, they often find themselves talking the same language – phase transitions, spontaneous symmetry-breaking, order parameters and low-lying collective modes. A phase transition is an instability leading to a macroscopic reorganization of a system's ground state, often involving spontaneous symmetry-breaking and the emergence of an order parameter. In condensed matter systems, such transitions are usually accompanied by major changes in bulk properties such as the specific heat and the magnetic susceptibility. Often the microscopic nature of the ordered phase can be characterized by direct external probes; for example the ordering of spins in a ferromagnet can be studied by application of a uniform magnetic field. However occasionally we are presented with a situation where the signatures of ordering are clearly present but the ordering microscopics remain elusive and inaccessible to simple observation. We must then do some detective work to glean more information. Often this is done by characterizing the resulting collective modes, with the hope of gaining insights about the underlying order that produced them. This approach is often taken in particle physics where the vacuum broken-symmetry state can only be probed in this fashion because we cannot put the Universe in an isolated box. Indeed the W and Z vector bosons are collective modes associated with the symmetry-breaking of the electroweak interaction; similarly the ongoing quest for the Higgs boson is motivated by the desire to characterize the collective modes of a cosmic order parameter.

The heavy electron metal $URu_2Si_2$ displays a classic second-order phase transition at $T_0 = 17.5$ K, yet the nature of the associated order parameter remains elusive more than two decades after its discovery.

This phase transition is characterized by a large entropy loss and sharp anomalies in a number of thermodynamic quantities (Fig. 1a).[1] Though this ordering was originally described in terms of spin density wave formation, this scenario had to be abandoned when neutron measurements revealed that the staggered magnetic moment was too small to account for the entropy loss at the transition.[2] At the time the presence of a propagating mode, attributed to a collective crystal-field excitation, was reported in the ordered phase[2] but its link to the nature of the underlying hidden order was far from clear. Fifteen years later an experimental reexamination of this collective mode by Wiebe et al.[3] on page 96 of this issue reveals new features that shed important light on the character of this hidden order.

$URu_2Si_2$ belongs to the family of heavy fermion materials where the conduction electrons develop large effective masses compared to those in conventional metals like copper. These heavy electrons live in a "twilight zone" between the extreme limits of localized and itinerant electronic behavior. From an electronic perspective, a heavy fermion metal is a lattice of localized magnetic moments immersed in mobile conduction electrons. At low temperatures local moments in these materials partially or completely dissolve into the surrounding conduction sea profoundly modifying their electronic properties. A signature of this dissolution is the emergence of heavy electrons with a marked proclivity towards exotic forms of ordering.

Part of the interest in $URu_2Si_2$ is due to the development of novel superconductivity in this material at low temperatures. Even the smallest amount of moment magnetism destroys pairing in conventional superconductors. A paradox of heavy electron systems is that they can develop superconductivity in the presence of dense lattices of local moments. Among the uranium heavy electron superconductors, there is a family of three materials ($UPd_2Al_3$, $UNi_2Al_3$ and $URu_2Si_2$) where there are two phase transitions where the lower one is superconducting. In the Al-based systems, these upper transitions involve the partial magnetic ordering of the local moments, and so it is natural to expect a similar situation in $URu_2Si_2$. However pressure studies indicate that the small-moment magnetism in this system is inhomogeneous and develops independently of the hidden order,[4] suggesting that a completely new type of electronic ordering is afoot.

Broadly speaking, there are two different theoretical views on the hidden order transition in the community. One group regards it as that of *localized* f-electrons on the uranium atoms. Since no significant vector spin order is observed, a multipolar (tensor) order parameter, allowed by the crystal structure, is proposed for the hidden order.[5] This local picture is able to reproduce the magnetic excitations that appear below $T_0$ and high-field phase behavior.[5] The second group regards the hidden order transition as an instability of the *itinerant* heavy fermions. The entropy loss is due to condensation of heavy fermion quasiparticles and is accompanied by the development of a gap in their excitation spectrum. Because conventional (isotropic) density waves are experimentally absent, anisotropic generalizations have been proposed[6,7] in analogy with exotic pairings in superconductivity. Even if one cannot detect the hidden order directly, collective modes should provide a way to discern between these two scenarios.

In a sequence of high-precision neutron scattering measurements, Wiebe et al.[3] have returned to the collective mode observed more than a decade ago.[2] These experiments have tracked the fate of the sharp collective mode previously observed in the hidden order phase as the temperature is raised through $T_0$. The surprising observation, contrary to previous expectations, is that the collective mode survives above $T_0$ as a damped and now gapless excitation (Fig 1b). The damping of this collective mode identifies it as a resonance in the particle-hole continuum of the metal (Fig 1c). This behavior is strongly reminiscent of paramagnetic fluctuations observed just above spin density wave transitions. Indeed borrowing from a model in spin density wave theory, the authors extract a correlation length for the underlying fluctuations which is several times larger than the unit cell at T= 20 K (Fig. 1c); they also find that the velocity of the quasiparticles contributing to this overdamped mode is in agreement with that of the heavy electrons. From a fit to their observed spectra, the authors also calculate the entropy of this mode and show that it agrees with that found from specific heat measurements. These observations all point towards the exciting possibility that the collective mode associated with the hidden order has been identified.

Still the mystery of the hidden order in $URu_2Si_2$ continues. The experiments of Wiebe *et al*[3] point to the clear importance of an itinerant picture: $URu_2Si_2$ is an anisotropically paired density wave. Many questions are raised. For example, the incommensurate nature of the collective mode strongly suggests that the heavy electron Fermi surface must contain "nested" surfaces, separated by the observed incommensurate wavevector. Another issue comes to mind. In $UPd_2Al_3$, a similar collective mode has been identified as the "glue" responsible for superconductivity.[8] It would be fascinating if an analogous situation occurs in $URu_2Si_2$, for then we may have a link between anisotropic pairing in hidden order and in exotic superconductivity.

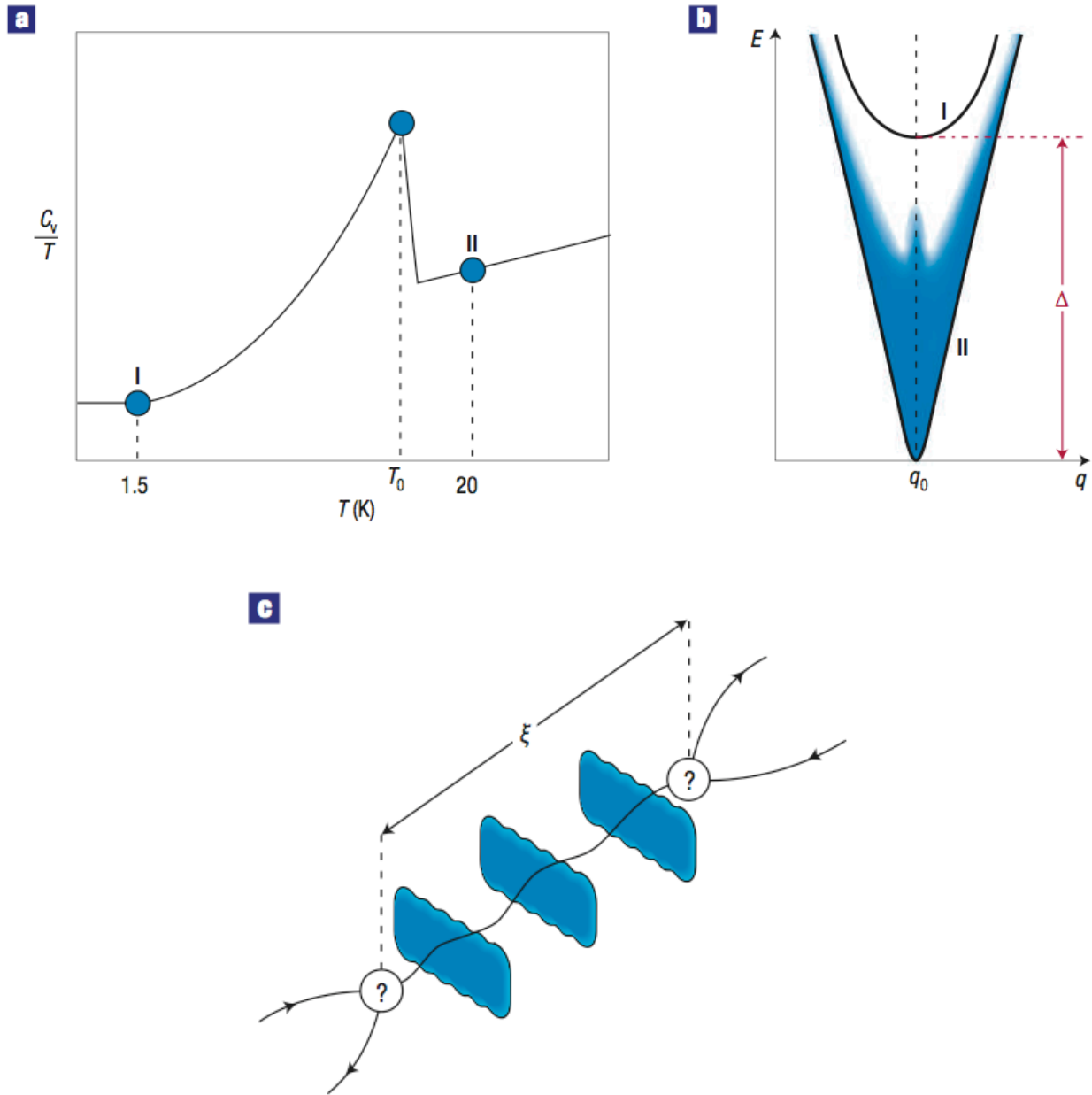

Figure 1: (Reproduced from reference [9])

What's hiding in $URu_2Si_2$? (a) Specific heat anomaly in $URu_2Si_2$, where I and II refer to temperatures where the spectrum of the collective mode has been measured. (b) Schematic showing the collective mode spectrum in the vicinity of the incommensurate wavevector (I) below $T_0$ showing the gap in the sharp collective mode and (II) above $T_0$ showing the gapless overdamped collective mode observed by Wiebe et al.[3] (c) Illustrating the decay of the collective mode into itinerant particle-hole pairs. The decay vertices, labelled by question marks, determine the anisotropic symmetry of the itinerant hidden order parameter and have yet to be determined. The correlation length $\xi$ determines the characteristic decay length of the two-body bound state.